\documentclass[aps,nofootinbib,preprintnumbers,superscriptaddress, prb]{revtex4-1}
\usepackage{bm}
\usepackage{hyperref}
\usepackage{amsmath,amssymb,amsfonts,graphicx,float}
\renewcommand{\vec}[1]{{\mathbf{#1}}}
\newcommand{\beq}{\begin{eqnarray}}
\newcommand{\eeq}{\end{eqnarray}}
\usepackage{amsmath}
\usepackage{amssymb}
\usepackage[paper=letterpaper,margin=1in]{geometry}
\usepackage{braket}
\usepackage{upgreek}
\usepackage{color}
\newcommand{\be}{\begin{equation}}
\newcommand{\ee}{\end{equation}}

\newcommand{\bwt}{\begin{widetext}}
\newcommand{\ewt}{\end{widetext}}

\def\a{\alpha}
\def\b{\beta}

\def\d{\delta}
\def\e{\epsilon}

\def\sig{\sigma}

\def\s{\sigma}

\begin{document}

\title{Non-equilibrium Transport in the Strange Metal and Pseudogap phases of the Cuprates}

\author{Ka Wai Lo}
\author{Seungmin Hong}
\author{Philip W. Phillips}
\affiliation{Department of Physics, University of Illinois at Urbana-Champaign, Urbana IL 61801, USA}

\begin{abstract}
We propose that the non-equilibrium current measured in the $a-b$ plane of an underdoped cuprate (in either the strange metal or pseudogap regime)  in contact with either an overdoped cuprate or a standard Fermi liquid can be used diagnose how different the pseudogap and strange metals are from a Fermi liquid.  Naively one expects the strange metal to be more different from a Fermi liquid than is the pseudogap.  We compute the expected non-equilibrium transport signal with the three Green functions that are available in the literature: 1) marginal Fermi liquid theory, 2) the phenomenological ansatz for the pseudogap regime and 3) the Wilsonian reduction of the Hubbard model which contains both the strange metal and pseudogap.  All three give linear IV curves at low bias voltages.  Significant deviations from linearity at higher voltages obtain only in the marginal Fermi liquid approach.  
The key finding, however, is that IV curves for the strange metal/Fermi liquid contact that exceed that of the pseudogap/Fermi liquid system.   
If this is borne out experimentally, this implies that the strange metal is less orthogonal to a Fermi liquid than is the pseudogap.  Within the Wilsonian reduction of the Hubbard model, this result is explained in terms of a composite-particle picture.  Namely, the pseudogap corresponds to a confinement transition of the charge degrees of freedom present in the strange metal.  In the strange metal the composite excitations break up and electron quasiparticles scatter off bosons.  The bosons here, however, do not arise from phonons but from the charge degrees of freedom responsible for dynamical spectral weight transfer.

\end{abstract}

\maketitle

\section{Introduction}

\begin{figure}
\begin{center}
\includegraphics[width=3.0in]{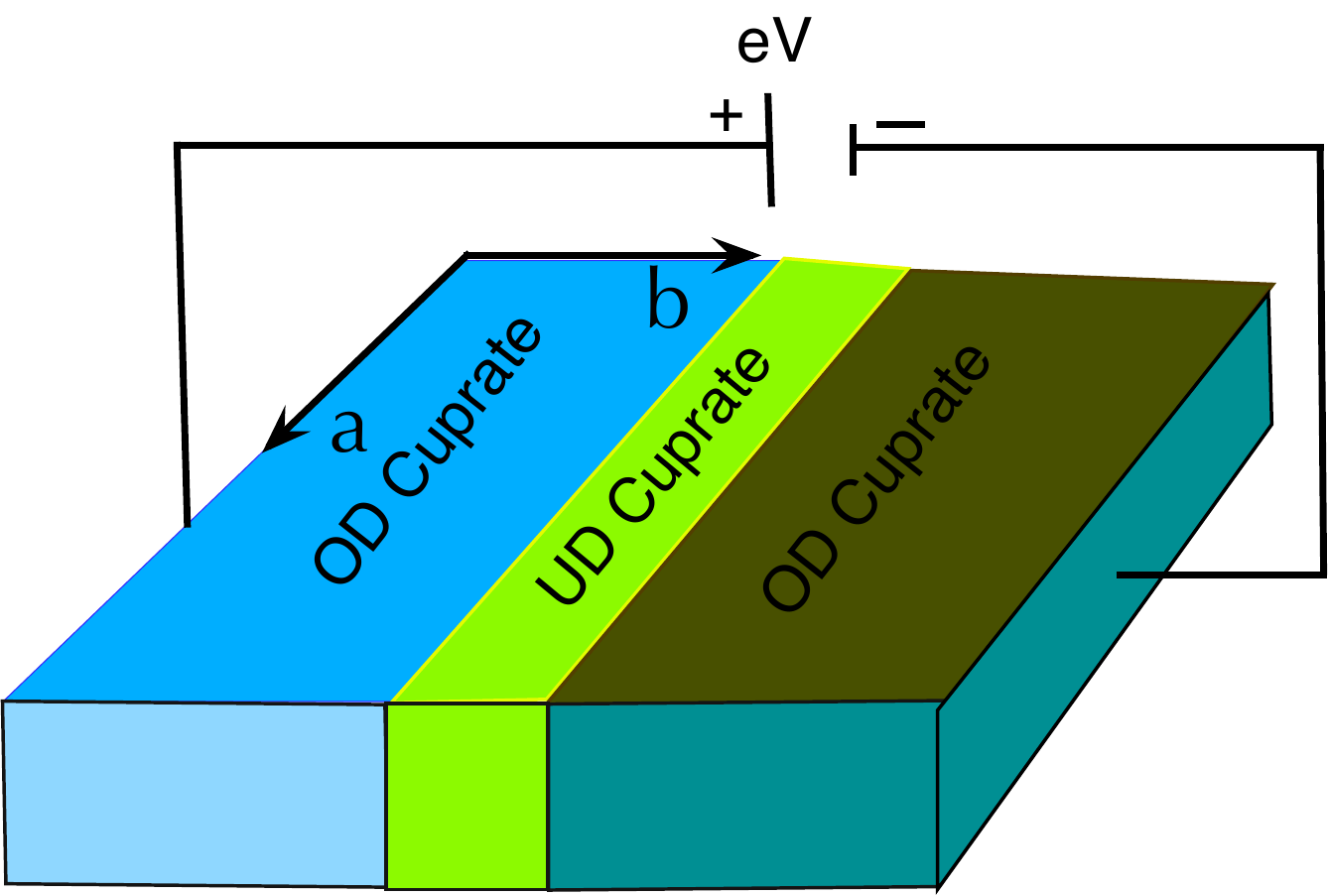}
\caption{  Device connecting an underdoped with an overdoped cuprate through a weak link.  The voltage difference is determined by the chemical potential difference between the two systems.  The overdoped cuprate functions as a reference Fermi liquid state and hence could be substituted with a metal. }
\label{geometry}
\end{center}
\end{figure}

While numerous experiments have been performed on the normal state of the cuprates, for example the littany of measurements now available on the pseudogap\cite{norman,alloul,npk,shekhter,lotfi,greven1,bourges,trsb1,trsb2,trsb3,trsb4,louis,nernst},  few have focused on the strange metal with an eye for identifying what is the nature of the current-carrying excitations that gives rise to $T-$linear\cite{ando, hussey} resistivity.   As a result, precisely what should go into  a theory of the normal state of the cuprates is missing. We propose here a simple experiment that can directly inform this problem.  Since the big problem is to understand precisely how each of these phases differs from a Fermi liquid, we propose that the geometry of interest is Fig. (\ref{geometry}) in which transport is measured in the $a-b$ plane of an underdoped cuprate in contact with either an overdoped material or a well known Fermi liquid metal.   Measuring the $I-V$ characteristics of an underdoped sample in this geometry in the pseudogap and the strange metal phases, separately accessible on a single sample simply by changing the temperature, will reveal how each differs from a Fermi liquid.  Since both phases exhibit features not in a Fermi liquid, they should each possess degrees of freedom orthogonal to electron quasiparticles which populate a Fermi liquid, such as an overdoped cuprate.  Since the strange metal possesses $T-$ linear resistivity, naively it is generally thought to be more different from a Fermi liquid than is the pseudogap.  This conclusion is certainly borne out by recent transport experiments\cite{greven}  on YBa$_2$Cu$_3$O$_{6+d}$ and YBa$_2$Cu$_4$O$_8$ which reveal the presence of a $T^2$ component in the resistivity in the pseudogap regime.  Hence, if the pseudogap has a noticeable Fermi liquid component, this would certainly affect effective modeling of this phase.  What we report here are a series of theoretical calculations of the I-V characteristics of Fig. (\ref{geometry}) using 1) marginal Fermi liquid theory\cite{mfl} for the strange metal, 2) a modified version of the phenomenological Green function of Yang, Rice and Zhang\cite{yrz} to model the pseudgap and 3) the composite-particle picture\cite{ftm1,ftm2,ftm3} obtained from a Wilsonian procedure on the Hubbard model for both the pseudogap and strange metals. These analyses point to the strange metal in contact with a Fermi liquid having a conductance that exceeds that of the pseudogap in the same geometry.  That is, our analysis indicates that the strange metal is less orthogonal to a Fermi liquid than is the pseudogap.  If this conclusion is borne out experimentally, then this would certainly place serious limitations on possible theories of the normal state.    

This paper is organized as follows.  In the next section, we describe the experimental setup and the non-equilibrium Green function formalism we adopt to calculate the conductance.  We describe in Section III the low-energy reduction of the Hubbard model and all the experimentally relevant results are included in Section IV.

\section{Non-equilibrium tunneling current}

The formalism to calculate the current through the device shown in Fig. (\ref{geometry}) can be obtained from the standard procedure\cite{Meir1992} for computing the conductance through a strongly correlated region attached to two non-interacting leads, each held at a different chemical potential\cite{Meir1992}.   The difference in chemical potential equals the applied voltage, $\mu_L-\mu_R = eV$. We consider the central region to be the strange metal or pseudo-gap phases of an underdoped cuprate, the only input required being the appropriate single-particle Green function.  In the next section specific models will be introduced for such Green functions.

The Hamiltonian of the full system consists of
\begin{align}\label{Hfull}
H=H_L+H_R+H_{c}+H_{\rm Hubb}
\end{align}
where $H_{Hubb}$ is the original Hubbard model which we write as
\begin{align}\label{Hubb}
H_{Hubb}=-t\sum_{ij} g_{ij} c^\dagger_{i\s}c_{j\s}+U\sum_{i\s} n_{i\uparrow}n_{i\downarrow}
\end{align}
with $g_{ij}=1$ for nearest neighbors and zero otherwise and $c_i$ is the electronic annihilation operator in the intermediate region at site $i$ ($i=1,\cdots,N)$ and $n_{i\s}$ is the density operator at site $i$ with spin $\s$, $n_{i\s}=c^\dagger_{i\s}c_{i\s}$..  The Hamiltonian for the leads is $H_{\b}=\sum_{k\s} \e_{\b,k} a^\dagger_{\b,k\s} a_{\b,k\s}$ ($\b=L,R$) which describes free fermions with a dispersion $\e_{\b,k}$ and creation (annihilation) operators $a^\dagger_{\b,k}$. The coupling term between the central region and the leads is of the form $H_{c}=\sum_{k i\s,\b} V^\b_{ki} a^\dagger_{\b,k\s} c_{i\s} + h.c. $, where $V^\b_{ki}$ is non-zero only if $i$ is neighboring  one of the leads.
This type of geometry has been studied previously  and the non-equilibrium tunneling current\cite{Meir1992,Jauho1994} is well known to be

\begin{align}\nonumber\label{J}
J&=\frac{ie}{2\hbar}\int \frac{d\e}{2\pi}{\rm Tr}\Big\{[\Gamma^L(\e)-\Gamma^R(\e)]G^<(\e)\\&+[f_L(\e) \Gamma^L(\e)-f_R(\e)\Gamma^R(\e)][G^r(\e)-G^a(\e)]\Big\}
\end{align}
where $[\Gamma^{\b}(\e_k)]_{ij}=2\pi\sum_{k\e \b}\rho_\b(\e_k)V_{ki}(\e_k)V^*_{kj}(\e_k)$, $\rho_\b$ is the density of states in the $\b$ lead,
$G^{r(a)}$ is the full retarded(advanced) Green function in the central region, 
$G^<$ is the full lesser Green function in the central region, which in the Schwinger-Kadanoff-Baym-Keldysh\cite{schwinger,baym,Keldysh1965} formalism is of the form
\beq\label{lesserG}
G^<=\mathcal{G}^<+\mathcal{G}^r\Sigma^{c,r}G^<
+\mathcal{G}^r\Sigma^{c,<}G^a
+\mathcal{G}^<\Sigma^{c,a}G^a.
\eeq
In these equations, $\mathcal{G}^{<,r,a}$ are the lesser, retarded and advanced Green functions for the central region {\it without} the coupling to the leads and hence are determined entirely from the low-energy physics of the central region.  These will be constructed in the next section.
$\Sigma^{c,<}$, $\Sigma^{c,r}$ and $\Sigma^{c,a}$ are the lesser, retarded and advanced self-energy contributions due to the coupling with the two leads.

In many cases, when the transport is dominated by states close to the Fermi energy, the wide band limit provides a good approximation to the retarded self energy \cite{Jauho1994}. In the wide band limit, $\Sigma^{c,r}=-i(\Gamma^L+\Gamma^R)/2$ where $\Gamma^L$ and $\Gamma^R$ are assumed to be independent of energy. 
Since the leads themselves are non-interacting, the lesser self-energy due to the coupling with the leads is simple and given by
$\Sigma^{c,<}_{ij}=\sum_{k,\b}V^\b_{ik}V^{\b,*}_{jk}g^{\b,<}(\e_k)$, where $g^{\b,<}$ is the lesser Green function for the two leads. In the wide band limit,
$\Sigma^{c,<}=i\Gamma^\b f_\b(\e)$ for the sites next to the $\b$ lead and equals zero otherwise.

To put all of this to use, we need expressions for Green functions that capture the physics of the central strongly correlated region.

\section{Infrared Theories of Central Region}

To model the central region, we consider 1) marginal-Fermi liquid phenomenology\cite{mfl} for the strange metal, 2) the standard mean-field ansatz\cite{yrz} for the pseudogap, and 3) the low-energy reduction of the Hubbard model obtained\cite{ftm1,ftm2,ftm3,shong} by formally integrating out the upper Hubbard band.   Only the 2nd and 3rd approaches require further enunciation.  

\subsection{Phenomenology}

Although not necessarily based on mean-field theory, the phenomenological form of the Green function proposed by Yang, Rice, and Zhang (hereafter YRZ)\cite{yrz}
\beq\label{RVB}
G^{\rm RVB}(\vec k,\omega)=\frac{g_t}{\omega-\xi(\vec k)-\frac{|\Delta(\vec
  k)|^2}{\omega+\epsilon^{\rm NN}(\vec k)}}+G_{\rm inc}
\eeq
is identical in form to that  proposed by ordering scenarios\cite{morr01,Norman2007,moon2010}, where $\epsilon^{\rm NN}(\vec k)=-2t(x)(\cos k_x+\cos k_y)$ (the
nearest-neighbour band structure) , $g_t=2x/(1+x)$, $\xi(\vec k)$
contains all the details of the band structure in the copper-oxide
plane including the chemical potential and $\Delta(\vec k)$ is of the
standard d-wave form.  The first term in this Green function can be written as a sum of two poles and hence can acquire a zero value.  It is the feature of zeros that has led to the popularity of the YRZ approach.   The stringent requirement for zeros of  the Green function is ${\rm Det }G(\omega,\vec p)=0$.  As a result a two-band non-interacting system which acquires the YRZ form when one of the bands is integrated out does not satisfy ${\rm Det}G(\omega,\vec p)=0$ as pointed out previously\cite{morr98,dave}.  Since the Hamiltonian can be diagonalized in a two-band non-interacting system, the determinant of the Green function is a product of resolvents (rather than a sum as would be the case in the Mott problem) and hence can never have  a true zero.  This is true of all mean-field models in which the underlying physics necessarily gives rise to a two-band system.  {\it A priori}, the YRZ Green function does not face this problem necessarily since no order is explicitly assumed.   In fact, the YRZ approach\cite{yrz}  is intended to model the effects of doping a multi-band ladder system (not a true 2D one) in which  the dopants leave one band effectively undoped at half-filling.   This is the origin of YRZ's placement of the zero line along the zone diagonal.  However, since this model has been used\cite{yrz} in the context of the cuprates, it is worth mentioning that a zero line remaining fixed along the zone diagonal, while relevant for a multi-band ladder system,  is not borne out by any simulations on the 2D Hubbard model\cite{kotstan,civelli11}.   Such a placement of the zero line guarantees that the particle density 
\beq\label{nlutt}
n=2\int_{\rm Re G(\omega,\vec p)>0} \frac{d^d\vec p}{(2\pi)^d},
\eeq
satisfies the Luttinger sum rule.  However, we have recently shown that any Green function which has zeros in the sense that  ${\rm Det} G(\omega=0,\vec p)=0$, Eq. (\ref{nlutt}) bares no relationship to the particle density.  The problem is that Eq. (\ref{nlutt}) implies a particle interpretation is valid for the charge density even when zeros of the Green function are present.  Zeros imply a divergent self energy at the chemical potential and hence no particle picture is present.  Precisely what replaces the Luttinger count when the self energy diverges is not known especially since  any statement of the form of Eq. (\ref{nlutt})  is possible only if an underlying Gaussian theory exists and no such Gaussian theory exists for zeros.

Nonetheless, it is worth comparing our results with the YRZ approach given its wide use.  The problem of how to compute the chemical potential in the YRZ approach arises immediately.  The standard procedure for fixing the chemical potential equates the filling
\beq\label{nlutt2}
n=\int_{-\infty}^\mu \sum_{\vec p} \Im G(\omega,\vec p) d\omega
\eeq
with an integral over the spectral function.  Note, Eqs. (\ref{nlutt}) and  (\ref{nlutt2}) are completely equivalent, since it is from the latter that the former is derived.  However, since the quantity $G_{\rm inc}$ is not known, the full spectral weight of the integration is determined entirely by the first term in Eq. (\ref{RVB}).  However,  as a result of the $g_t$ factor in the numerator in Eq. (\ref{RVB}), a quantity that  vanishes at $x=0$, this expression can never yield a filling of unity as is required at half-filling.  Table (\ref{chempot}) illustrates the problem.  For $x<0.25$, the  upper bound on the doping level for which the YRZ ansatz is expected to hold, there is no solution to Eq. (\ref{RVB}) for the chemical potential.  Only for $x>0.4$ do we obtain a chemical potential of the correct sign when the coherence factor $g_t$ is retained.  Since this is outside the range of validity of the YRZ scheme, we considered, as did others\cite{tremblay}, the case in which $g_t=1$.   As is evident from Table (\ref{chempot}), reasonable results obtain which are identical with those of YRZ\cite{yrz}.  However, this procedure is not entirely consistent.  Certainly, if the first term in Eq. (\ref{RVB}) described just a Fermi liquid, then grouping any such multiplicative factors into the incoherent part would be justified.  However, the novel physics that Eq. (\ref{RVB}) is designed to capture is that of zeros for which no coherent quasiparticle exists.   Nonetheless, for the sake of comparison we adopt the procedure in which the first term in Eq. (\ref{RVB}) is divided by $g_t$.   We refer to all results obtained in this fashion as $YRZ/g_t$.

\begin{table}[!h] 
\tabcolsep 0pt \caption{Comparison of the chemical potential values in the YRZ phenomenological approach with
and without the factor of $g_t=2x/(1+x)$ in the first term in Eq. (\ref{RVB}).  For a hole-doped system, the chemical potential should be negative.  N. A. stands for not available as there is either no solution, $x<0.25$ with $g_t$ present or the parameter range is beyond the limit of validity of the YRZ ansatz, $x>0.25$.  The correct sign for the chemical potential obtains in the doping range of interest, $x<0.25$, only when $g_t=1$.  }

\begin{center}

\def\temptablewidth{0.47\textwidth}

\begin{tabular*}{\temptablewidth}{@{\extracolsep{\fill}}|c|c|c|c|}

\hline {\rm doping level (x) } & $g_t$& $\mu (g_t)$& $\mu(g_t=1)$\\
\hline
0.05& 0.09&{\rm N. A.}&-0.091\\
\hline 0.08&0.15&{\rm N. A.}&-0.161\\
\hline 0.12&0.21&{\rm N. A.} &-0.25\\
\hline 0.15&0.26&{\rm N. A.}&-0.315\\
\hline 0.18&0.31&{\rm N. A.}&-0.384\\
\hline 0.25&0.40&2.029&{\rm N. A.}\\
\hline 0.30&0.46&0.9622&{\rm N. A.}\\
\hline 0.35&0.52&0.254&{\rm N. A.}\\
\hline 0.40&0.57&-0.242&{\rm N. A.}\\
\hline

\end{tabular*}
\end{center} \label{chempot}
\end{table}

\subsection{Wilsonian Procedure}

Integrating out the upper Hubbard band can be done exactly to yield a single theory that contains the strange metal at high temperatures and a pseudogap at low temperatures.  To review the details of this derivation, we consider a slightly different starting point of the Hubbard model
\begin{align}\label{H}
H &=\sum_{\langle ij\rangle}-t_{ij}c^\dagger_{i,\s}c_{j,\s}-\sum_{i,\s} \mu_i n_{i\s} +U\sum_in_{i\uparrow}n_{i\downarrow}\nonumber \\
 &+\left(\sum_{i,k_P,\s,P=L,R} V_{k_P,i,\s}c_{i,\s}^\dagger a_{k_P,\s}+h.c.\right)\nonumber \\
 &+\sum_{k_P,\s,P=L,R} \e_{k_P,\s} n_{k_P,\s}
\end{align}
which incorporates the leads explicitly where $a_{{k_P},\s}$ annihilates an electron, the index $P=L,R$ represents the leads with momentum $k_P$ and spin $\s$. $V_{k_P,i,\s}$ describes the tunneling rate between electrons at site $i$ in the intermediate region and the electron at the $P$ lead with momentum $k_P$. We will assume it is non-vanishing only when $i=1$ or $N$. 

Since the intermediate region is described by the Hubbard model in the strongly correlated regime, mean-field decoupling is insufficient \cite{Yonemitsu2009}. What should be done then is correctly integrate out the degrees of freedom far way from the chemical potential, which for a hole-doped system is the upper Hubbard band.  As these bands are not static, this is non-trivial because dynamically generated degrees of freedom will appear at low energy.  What we have done is introduce a new fermionic degree of freedom which creates excitations at energy scale $U$.  Only when it is constrained does it correspond to double occupancy.  This is important because only in the atomic limit of the Hubbard model does can the eigenstates be indexed according to energy by the number of double occupancies that are present.  The correct starting point is
\begin{widetext}
\begin{align}\nonumber
L=&\int d^2\theta \Big[\bar{\theta}\theta s\sum_{i,\sigma}(1-n_{i,\bar{\sigma}}) c^\dagger_{i\sigma}\dot{c}_{i\s}-\sum_{i,\s} \mu_i n_{i\s}+\sum_i D^\dagger_i\dot{D}_i+U\sum_j D^\dagger_j D_j
\\\nonumber&-t\sum_{ij\s}g_{ij}[C_{ij\s}c^\dagger_{i\s}c_{j\s}+D^\dagger_i c^\dagger_{j\s}c_{i\s} D_j+(D^\dagger_j\theta c_{i\s}V_\s c_{j\bar{\s}}+h.c)]
\\&+\sum_{i,k_P,\s}\Big(V_{k_P,i\s}a_{k_P,\s}\big(\bar{\theta}\theta(1-n_{i,\s})c^\dagger_{i\s}-\theta c_{i,\bar{\s}}V_\s D^\dagger_i\big)+h.c.\Big)+\bar{\theta}\theta \sum_{k_P,\s}\e_{k_P,\s}n_{k_P,\s}+H_{con}+H_J\Big]
\end{align}
\end{widetext}
where the constraint is
\begin{align}
H_{con}=s\bar{\theta}\sum_j \varphi_j^\dagger(D_j-\theta c_{j,\uparrow}c_{j,\downarrow})+h.c.,
\end{align}
and the source term for the charge is
\beq
H_J=J_{i\sig}^*\left[\bar\theta\theta(1-n_{i\bar\sig})c_{i\sig}
	+ \bar\theta c_{i\bar\sig}^*V_\sig D_i\right]+\text{c.c.}
\eeq
with $\theta$ a Grassman variable, $V_\uparrow=-V_\downarrow=1$ and $\varphi$ is a bosonic field that enters the theory as a Lagrange multiplier.   Each of these terms is designed so that when the constraint is solved in the standard Fermionic path integral formulation, the starting Hubbard model results.  In this limit, the source term reduces to $J_\sig c_{i\sig}$, namely the UV charge.  However, the form of this Lagrangian permits an explicit integration of the physics on the $U-$scale simply by integrating out the $D_i$ fields.   The result is the true-low energy physics
\begin{widetext}
\begin{align}\label{HIR}
H_{IR}=&-t\sum_{i,j,\s}g_{ij}\a_{ij,\s}c^\dagger_{i,\s}c_{j,\s}-\sum_{i,\s} (1-n_{i\bar{\s}})\mu_i n_{i\s}(1-n_{i\bar{\s}})+ H_{int}-\frac{1}{\b}{\rm Tr}\ln \mathcal{M}\\\nonumber
&+(\sum_{k_P,i,\s,P} V_{k_P,i\s}\psi^\dagger_{i\s} a_{k_P} +h.c.)-\sum_{ij\s}\sum_{k_P,P} V^*_{k_P,i,\s} a_{k_P,\s}^\dagger c^\dagger_{i\bar{\s}} (\mathcal{M}^{-1})_{ij} c_{j\bar{\s}} \sum_{k'_{P'},P'}V_{k'_{P'},j,\s}a_{k'_{P'}\s}\\
&+\sum_{k_P,\s,P}\e_{k_P\s}n_{k_P,\s}
\end{align}
\end{widetext}
in the presence of the leads.  Here       
\begin{align}
H_{int}&=(s\varphi_i-tb_i)^\dagger(\mathcal{M}^{-1})(s\varphi_j-tb_j)\nonumber\\
	&\qquad -(s\varphi_i^\dagger c_{i\uparrow}c_{i\downarrow}+h.c.)\\
\mathcal{M}_{ij}&=\left[(\omega-\mu_i+U)\d_{ij}-tg_{ij}\sum_\s c^\dagger_{j,\s}c_{i,\s}\right]
\end{align}
with $b_i=\sum_j g_{ij}c_{j\s}V_\s c_{i\bar{\s}}$. The presence of the Lagrange multiplier, $\varphi_i$ in the low-energy theory is the generator of dynamical spectral weight transfer in the lower band.  
The last term in the second line of $\eqref{HIR}$ describes a second-order hopping process between the lead and central region.  Also of interest is the form the source term  acquires
\beq
\psi_{i\sig}^* &=& (1-n_{i\bar\sig})c_{i\sig}^* +tb_j^*\left({ M}^{-1}\right)_{ji}V_{\sig}c_{i\bar\sig}\nonumber\\
	&&\quad-s\varphi_j^*\left( M^{-1}\right)_{ji}V_{\sig}c_{i\bar\sig}.
\eeq
upon integration over $D_i$.  This expression will be our working equation for the IR excitations in the central region.  The first term in the expression is the standard projected electron in the lower band, the second arising from spin fluctuations, and the third the new degree of freedom associated with the dynamical transfer of spectral weight from the upper band.  Since such degrees of freedom have internal degrees of freedom as a result of the $\varphi$ dependence, they are orthogonal to the projected part of the electron operator and hence can only give rise to zeros of the electron propagator.  We showed this more precisely by deriving the Green function\cite{shong}
\begin{widetext}
\begin{align}\label{2eBG}
\mathcal{G}(i\omega_n,\bold{k})&=\langle T_\tau \psi_i(\tau) \psi^\dagger_j(0)\rangle= \frac{\tilde{g_t}}{i\omega_n-\mu-\tilde{g_t}\epsilon_\bold{k}-\Sigma_\pm(i\omega_n,\bold{k})}+O(\frac{t}{U}),\\
\Sigma_\pm(i\omega_n,\bold{k}) &= \frac{s_{\bold{k},\bold{q}}^2\varphi_q\varphi_q^*} {i\omega_n-\mu\pm \tilde{g_t}\epsilon_{\bold{q}-\bold{k}}},
\end{align}
\end{widetext}
for the IR degrees of freedom.  Here $\tilde{g_t}=g_t/g_p$, $g_t = 2(x+\a)/(1+x)$ and $g_p = (1-x-\a)/(1-x)$ is the projection factor introduced in \cite{Zhang1988}.
The the $\pm$ sign in the denominator represents two different ways to treat the dynamics of charge 2e field $\varphi$. The $+$ sign is used when the $\varphi$ field is treated as an independent degree of freedom that can condense in particle-particle channel.  This corresponds to the strange metal regime.  The $-$ sign treats $\varphi^*_i c_{i\bar{\s}}$ as a bound object resulted from dynamical spectral weight transfer; hence the $\varphi^*_i c_{i\s} c_{i\bar{\s}}$ term in $H_{int}$ resembles particle-hole channel scattering event, thereby creating the pseudogap with a Fermi arc structure reminiscent of what is reported in  ARPES on the normal state of the underdoped cuprates.  Our expressions here for $\mathcal{G}$ will form the basis for an analysis of the tunneling current in Fig. (\ref{geometry}).  On physical grounds because the transition from the strange metal to the pseudogap is one of confinement, it is the strange metal that should have the higher tunneling current into the adjacent Fermi liquid in the geometry shown in Fig. (\ref{geometry}).  Our calculations will show this explicitly.

From the effective low-energy Hamiltonian $H_{IR}$, we can derive the current flow through the intermediate region readily. 
\begin{align}\nonumber
J_L&=-e\langle \dot{N}_L\rangle=-\frac{ie}{\hbar}\langle [H_{IR},N_L]\rangle\\
&=\frac{ie}{\hbar} \sum_{k_L,i\s}[V^*_{k_L,i,\s}\langle a^\dagger_{k_L,\s}\psi_{i\s}\rangle- V_{k_L,i,\s}\langle \psi^\dagger_{i\s}a_{k_L,\s}\rangle]
\end{align}
The term $\langle a^\dagger_{k_L,\sigma,\s}\psi_{i\s}\rangle$ and its hermitian conjugate can be computed using the equation of motion approach following \cite{Meir1992}. We note that
\begin{align}
\dot{a}^\dagger_{k_P,\s}&=i[H_{IR},a^\dagger_{k_P,\s}]\\
&=i\e_{k_P,\s}a^\dagger_{k_P,\s}+i\sum_{i\s} V_{k_P,i\s}\psi^\dagger_{i\s}-i\sum_{ij}\sum_{k'_{P'}P'}V^*_{k'_{P'}i\s} a^\dagger_{k'_{P'}} c^\dagger_{i\bar{\s}} (\mathcal{M}^{-1})_{ij} c_{j\bar{\s}} V_{k_Pj\s}.\label{aeom}
\end{align}
 Keeping only the leading order term, we find that
\begin{align}
-i\frac{\partial}{\partial t}G^t_{k_L,i,\s}(t-t')=\e_{k_P}G^t_{k_L,i,\s}(t-t')+\sum_{j} V_{k_L,j,\s} G^t_{ji,\s}(t-t')
\end{align}
where
\begin{align}
G^t_{k_L,i\s}(t-t')=-i\langle T\{ a^\dagger_{k_L,\s}(t')\psi_{i\s}(t)\}\rangle\\
G^t_{j,i\s}(t-t')=-i\langle T\{ \psi^\dagger_{j,\s}(t')\psi_{i\s}(t)\}\rangle.
\end{align}
The current is then given by
\begin{align}\label{current}
J_L=\frac{2e}{\hbar}\int \frac{d\e}{2\pi} {\rm Re}\Big\{ \sum_{k'_L,k_L,i,j,\s} V_{k'_L,i\s}^*V_{k_L,j,\s} [g^r_{k'_L,k_L,\s}(\e)G^<_{ji,\s}(\e)+g^<_{k'_L,k_L,\s}(\e)G^a_{ji,\s}(\e)]\Big\}
\end{align}
where
$g^{t-1}_{k_L,k'_L}=(-i\partial/\partial t-\e_k)\d_{kk'}$.
Hence, we have recovered Eq.\eqref{J} except that the Green function in the central region is the propagator of $\psi$ instead of the bare electron.

In Eq. \eqref{current}, $G^a$ can be obtained from Eq. \eqref{2eBG} by including the self-energy contribution from coupling to the leads $\Sigma^{c,a}$. In order to evaluate $G^<$ which is given by Eq.\eqref{lesserG},
we will also need the lesser Green function in the central region given by $\mathcal{G}^< = \mathcal{G}^r \Sigma^{\varphi,<} \mathcal{G}^a$. $\Sigma^{\varphi,<}$ is the lesser self-energy due to the charge 2e boson field and is given by
\beq
\Sigma^{\varphi,<}_{ij}=s^2\varphi_i \varphi_j^*G_{ij}^{0,<},
\eeq
and $G_{ij}^{0,<}$ is the lesser hole propagator for the tight-binding part of the Hamiltonian in \eqref{HIR} only.

\subsection{Results}

The work outlined thus far formulates the current in real space.  However,  neither the MFL\cite{mfl} nor the Green functions indicative of mean-field theory\cite{yrz} have real-space analogues.  Consequently, to make contact with this work, we must resort to momentum space and hence we use a particular ansatz
\beq
\Sigma^{<} &=&\Sigma^{<}_0\left(1-\frac{2{\rm Im}\Sigma^r}{\Gamma}\right)\\
\Sigma^<_0&=&i\left[f_L(\e)\Gamma_L\left(\e+\frac{eV}{2}\right)+f_R(\e)\Gamma_R\left(\e-\frac{eV}{2}\right)\right)\nonumber \\
\eeq
 for the lesser self energy proposed previously\cite{Ng1996}, where $\Sigma^r$ is the retarded self-energy of the Green function in the central region when it is decoupled from the leads. 
This phenomenological calculation allows us to compute the non-equilibrium tunneling current for any Green function once $\Sigma^r$ is known. Hence, we can compare the result from the charge 2e boson theory with other Green functions that have been used to model the pseudo-gap phase and strange metal phase of underdoped cuprate.

Nonetheless, this approaches is not strictly correct for the charge 2e boson theory as it assumes the full lesser Green function can be written as $G^<=G^r\Sigma^<G^a$. This is not the case for the charge 2e boson as the two components of the lesser self-energy, $\Sigma^{c,<}$ and $\Sigma^{\varphi,<}$ can not be added directly due to the projection factor $g_t$. In order to determine $G^<$ correctly, we must use Eq. \eqref{lesserG} which we will do in the real space calculation in the next subsection. 
Nevertheless, as we will see, the result from phenomenological calculation matches the results from real space calculation qualitatively.

\begin{figure*}[!ht]
\centering
\hskip -0.09in\includegraphics[height=5cm]{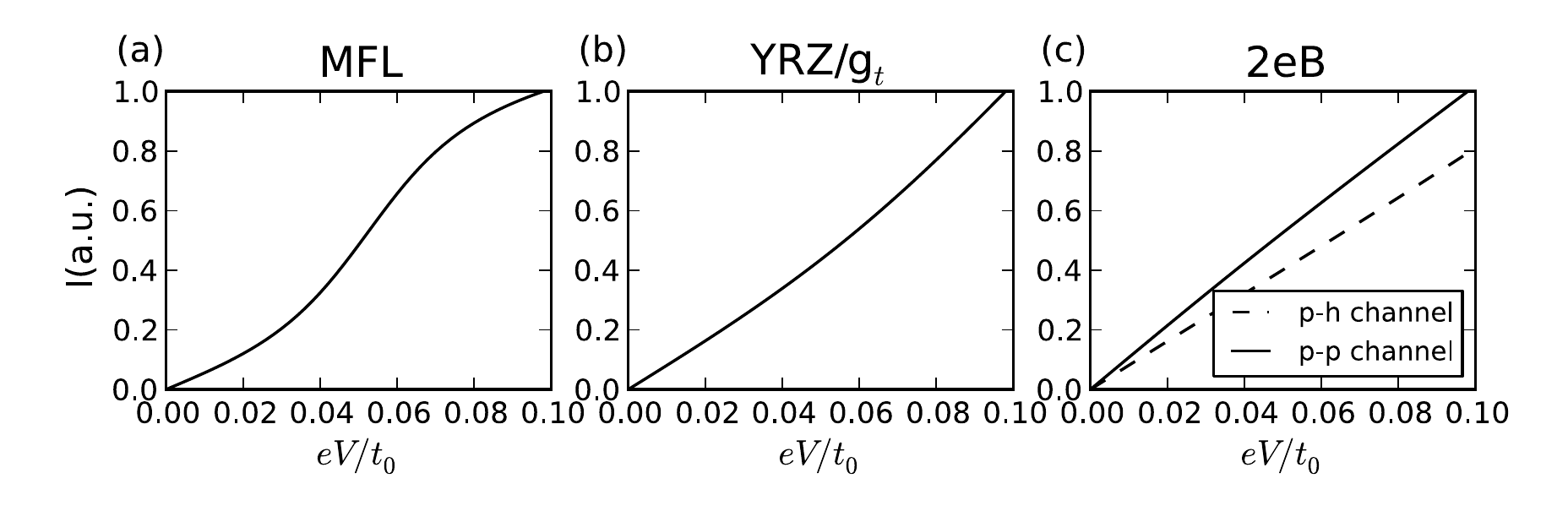}
\caption{\footnotesize{\label{IVcurve}The non-equilibrium current as a function of applied voltage for three different Green functions. The leftmost one corresponds to Marginal Fermi liquid. The middle one corresponds to the results of the YRZ Green function divided by the factor $g_t=2x/(1+x)$, while the rightmost one corresponds to the charge 2e boson theory.} }
\end{figure*}

Figure \ref{IVcurve} shows the (normalized) non-equilibrium $I-V$ characteristic curve for a MFL, YRZ Green functions and charge 2e boson for $\Gamma_L=\Gamma_R=0.01t_0$ respectively.  The $I-V$ characteristic curves are linear when the applied voltage is small for all three Green functions. As the voltage increases, the marginal Fermi liquid first deviates from and increases more rapidly from  linear and as the voltage continues to increase, the rate of increase of the current (differential conductance) drops again and becoming approximately constant as $eV\sim 0.1 t_0$.

For the YRZ Green function, the $I-V$ characteristic curve is quite linear throughout the range of applied voltage of interest here. It only deviates from linear behavior slightly when $eV\sim 0.05 t_0$. For the charge 2e boson theory, the $I-V$ characteristic behaves even more linear compared to the YRZ Green function. It does not show significant deviation from linear behavior throughout the range of applied bias voltage in consideration. The linear behavior is observed for both the particle-particle (strange metal) and particle-hole (pseudogap) channels. However, the current is larger in the particle-particle channel than the particle-hole channel for the same doping (x=0.12) as shown in figure \ref{IVcurve}(c).
As mentioned previously, this is expected within the framework of the transition from the strange metal to the strange metal being one of confinement.  That is, in the strange metal, the charge 2e boson does not mediate new degrees of freedom but simply acts as a local scatterer from projected electrons.  In terms of the UV variables, this corresponds to freely moving doublons and holons.  They bind at low temperatures thereby giving rise to a decrease in the differential conductance and thereby the pseudogap.  While it is not possible to determine the scale for the MFL and YRZ Green functions, the MFL result certainly exceeds the tunneling characteristics in the pseudogap phase at high bias voltages.  

There is a serious limitation in these phenomenological calculations in that they do not capture the variation in the local chemical potential and number density due to the applied bias voltage. These effects are only negligible when the applied voltage is small. In order to consider this effect, we must compute the non-equilibrium current  in real space and the underlying microscopic theory for the the central interacting region must be known. As a result, the phenomenological YRZ and marginal fermi liquid Green function cannot be used, and we are faced to resort only to to the charge $2e$ theory.   Similar theory accounts such as the co-fermion account\cite{imada} should be in qualitative agreement with the current work as also in this work, the Fermi arc structure is attributed  to bound state formation.

\begin{figure}[!b]
\hskip -0.19in\includegraphics[width=1.1\columnwidth]{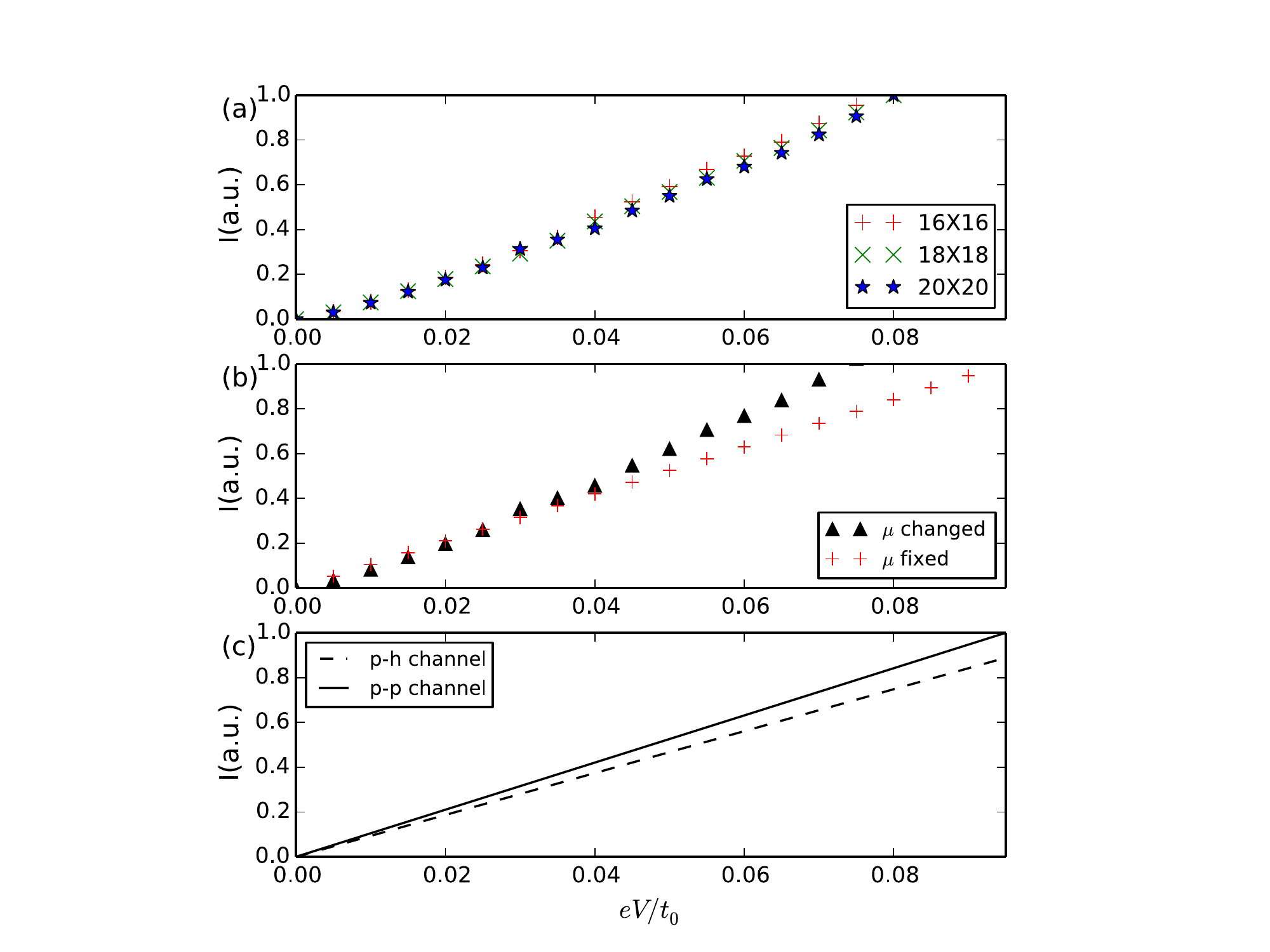}
\caption{\footnotesize{\label{IVreal}(a) The normalized
    non-equilibrium current as a function of applied bias voltage for
    square lattice size of $16\times16$, $18\times18$ and $20\times20$
    in the central interacting region in particle-hole channel.  The
    variation of the chemical potential is taken into consideration by
    Eq. (\ref{muchange}). 
(b) The tunneling current for square lattice size of $16\times16$, the black triangle corresponds to the case where the local chemical potential and number density is modified due to the applied bias voltage. The red plus sign corresponds to the case where the chemical potential is kept fixed.
(c) The non-equilibrium tunneling current for particle-particle channel (solid line) and particle-hole channel (dashed-line). The background chemical potential is the same for both cases.} }
\end{figure}

Now for the variation of the chemical potential.  Working in real space allows us to consider the local change in the chemical potential in terms of the doping level. For simplicity, we will assume a simple linear interpolation between the chemical potential at left and right leads:
\beq\label{muchange}
\mu(x,y) =
\begin{cases}
\mu_0+e V/2, & \text{for }x\leq1 \\
\mu_0 + eV/2 -eV \frac{x-1}{L_x-1} & \text{for } 1<x<Lx \qquad \\ 
\mu0-e V/2, & \text{for }x\geq L_x.
\end{cases}
\eeq

Including the variation in the chemical potential can lead to a
system size dependence.  To calibrate how our results depend on system size, we demonstrate the size
dependence in Fig. (\ref{IVreal}) (a).  The current is normalized by its value at $eV= 0.80 t_0$. As is evident, by a lattice of
size of $20\times20$, all size dependence vanishes.  Consequently,
all our subsequent calculations were performed on a lattice size of
$20\times 20$. 
Fig. \ref{IVreal}(b) shows the effect on the $I-V$ curves due to the modification of the chemical potential according to Eq.\eqref{muchange} (black triangle). We  see that the $I-V$ curve is again roughly linear for small applied bias voltage though it starts to deviate from linear behavior when $eV\sim 0.02 t_0$. Note that we cannot go further away from $eV = 0.08t_0$ because the change in the chemical potential is so large that the corresponding doping range exceeds that of the pseudogap regime and the charge $2e$ formalism no longer applies.   We also consider the difference in tunneling current if we decouple the interaction term $\varphi^*_i c_{i\s}c_{i\bar{\s}}$ in particle-hole and particle-particle channels. Figure \ref{IVreal}(c) shows the result where the background chemical potential is the same for both cases. We again see that the current is larger in the particle-particle channel than the particle-hole channel, justifying our result in the phenomenological calculation and our intuition

\section{Conclusion}

We have proposed here that the device in Fig. (\ref{geometry}) can be used to infer how different the pseudogap and strange metals are from a Fermi liquid.  Surprisingly,
we find that the conductance between an underdoped cuprate and a metal  exceeds that in the pseudogap regime.  In addition, our results indicate that the IV curves should exhibit a considerable range where they are linear.  Should this trend be borne out experimentally, serious 
constraints will then be placed on candidate explanations for the strange metal as it suggests that the strange metal has the same carriers 
as does a normal metal.  What makes the strange metal strange is that it possesses an additional bosonic excitation which acts as a scattering center for the electronic degrees of freedom which emerges at low energies from explicitly integrating out the upper Hubbard band.

\section*{Acknowledgements}
We thank I. Bozovic and C. Panagopoulos for useful conversations at the inception of this project.   The S. Hong and P. Phillips is supported by NSF DMR- and the
the work of Ka Wai Lo is funded through the Center for Emergent Superconductivity, a DOE Energy Frontier Research Center, Grant No.~DE-AC0298CH1088. 

\begin{thebibliography}{10}%
\makeatletter
\providecommand \@ifxundefined [1]{%
 \ifx #1\undefined \expandafter \@firstoftwo
 \else \expandafter \@secondoftwo
\fi
}%
\providecommand \@ifnum [1]{%
 \ifnum #1\expandafter \@firstoftwo
 \else \expandafter \@secondoftwo
\fi
}%
\providecommand \enquote [1]{``#1''}%
\providecommand \bibnamefont  [1]{#1}%
\providecommand \bibfnamefont [1]{#1}%
\providecommand \citenamefont [1]{#1}%
\providecommand\href[0]{\@sanitize\@href}%
\providecommand\@href[1]{\endgroup\@@startlink{#1}\endgroup\@@href}%
\providecommand\@@href[1]{#1\@@endlink}%
\providecommand \@sanitize [0]{\begingroup\catcode`\&12\catcode`\#12\relax}%
\@ifxundefined \pdfoutput {\@firstoftwo}{%
 \@ifnum{\z@=\pdfoutput}{\@firstoftwo}{\@secondoftwo}%
}{%
 \providecommand\@@startlink[1]{\leavevmode}%
 \providecommand\@@endlink[0]{}%
}{%
 \providecommand\@@startlink[1]{%
  \leavevmode
  \pdfstartlink
   attr{/Border[0 0 1 ]/H/I/C[0 1 1]}%
   user{/Subtype/Link/A<</Type/Action/S/URI/URI(#1)>>}%
  \relax
 }%
 \providecommand\@@endlink[0]{\pdfendlink}%
}%
\providecommand \url  [0]{\begingroup\@sanitize \@url }%
\providecommand \@url [1]{\endgroup\@href {#1}{\urlprefix}}%
\providecommand \urlprefix [0]{URL }%
\providecommand \Eprint[0]{\href }%
\@ifxundefined \urlstyle {%
  \providecommand \doi [1]{doi:\discretionary{}{}{}#1}%
}{%
  \providecommand \doi [0]{doi:\discretionary{}{}{}\begingroup
  \urlstyle{rm}\Url }%
}%
\providecommand \doibase [0]{http://dx.doi.org/}%
\providecommand \Doi[1]{\href{\doibase#1}}%
\providecommand \bibAnnote [3]{%
  \BibitemShut{#1}%
  \begin{quotation}\noindent
    \textsc{Key:}\ #2\\\textsc{Annotation:}\ #3%
  \end{quotation}%
}%
\providecommand \bibAnnoteFile [2]{%
  \IfFileExists{#2}{\bibAnnote {#1} {#2} {\input{#2}}}{}%
}%
\providecommand \typeout [0]{\immediate \write \m@ne }%
\providecommand \selectlanguage [0]{\@gobble}%
\providecommand \bibinfo [0]{\@secondoftwo}%
\providecommand \bibfield [0]{\@secondoftwo}%
\providecommand \translation [1]{[#1]}%
\providecommand \BibitemOpen[0]{}%
\providecommand \bibitemStop [0]{}%
\providecommand \bibitemNoStop [0]{.\EOS\space}%
\providecommand \EOS [0]{\spacefactor3000\relax}%
\providecommand \BibitemShut [1]{\csname bibitem#1\endcsname}%
\bibitem{norman}%
  \BibitemOpen
  \bibfield{author}{%
  \bibinfo {author} {\bibfnamefont{M.~R.}\ \bibnamefont{Norman}}, \bibinfo
  {author} {\bibfnamefont{H.}~\bibnamefont{Ding}}, \bibinfo {author}
  {\bibfnamefont{M.}~\bibnamefont{Randeria}}, \bibinfo {author}
  {\bibfnamefont{J.~C.}\ \bibnamefont{Campuzano}}, \bibinfo {author}
  {\bibfnamefont{T.}~\bibnamefont{Yokoya}}, \bibinfo {author}
  {\bibfnamefont{T.}~\bibnamefont{Takeuchi}}, \bibinfo {author}
  {\bibfnamefont{T.}~\bibnamefont{Takahashi}}, \bibinfo {author}
  {\bibfnamefont{T.}~\bibnamefont{Mochiku}}, \bibinfo {author}
  {\bibfnamefont{K.}~\bibnamefont{Kadowaki}}, \bibinfo {author}
  {\bibfnamefont{P.}~\bibnamefont{Guptasarma}},\ and\ \bibinfo {author}
  {\bibfnamefont{D.~G.}\ \bibnamefont{Hinks}},\ }%
  \bibfield{journal}{%
  \bibinfo {journal} {Nature}\ }%
  \textbf{\bibinfo {volume} {392}},\ \bibinfo {pages} {157} (\bibinfo {year}
  {1998}),\ \url{http://dx.doi.org/10.1038/32366}%
  \bibAnnote{NoStop}{norman}{10.1038/32366.}%
\bibitem{alloul}%
  \BibitemOpen
  \bibfield{author}{%
  \bibinfo {author} {\bibfnamefont{H.}~\bibnamefont{Alloul}}, \bibinfo {author}
  {\bibfnamefont{T.}~\bibnamefont{Ohno}},\ and\ \bibinfo {author}
  {\bibfnamefont{P.}~\bibnamefont{Mendels}},\ }%
  \bibfield{journal}{%
  \Doi{10.1103/PhysRevLett.63.1700}{\bibinfo {journal} {Phys. Rev. Lett.}}\ }%
  \textbf{\bibinfo {volume} {63}},\ \bibinfo {pages} {1700} (\bibinfo {month}
  {Oct}\ \bibinfo {year} {1989}),\
  \url{http://link.aps.org/doi/10.1103/PhysRevLett.63.1700}%
  \bibAnnoteFile{NoStop}{alloul}%
\bibitem{npk}%
  \BibitemOpen
  \bibfield{author}{%
  \bibinfo {author} {\bibfnamefont{M.~R.}\ \bibnamefont{Norman}}, \bibinfo
  {author} {\bibfnamefont{D.}~\bibnamefont{Pines}},\ and\ \bibinfo {author}
  {\bibfnamefont{C.}~\bibnamefont{Kallin}},\ }%
  \bibfield{journal}{%
  \bibinfo {journal} {Advances in Physics}\ }%
  \textbf{\bibinfo {volume} {54}},\ \bibinfo {pages} {715} (\bibinfo {year}
  {2005})%
  \bibAnnoteFile{NoStop}{npk}%
\bibitem{shekhter}%
  \BibitemOpen
  \bibfield{author}{%
  \bibinfo {author} {\bibfnamefont{A.}~\bibnamefont{Shekhter}}, \bibinfo
  {author} {\bibfnamefont{B.~J.}\ \bibnamefont{Ramshaw}}, \bibinfo {author}
  {\bibfnamefont{R.}~\bibnamefont{Liang}}, \bibinfo {author}
  {\bibfnamefont{W.~N.}\ \bibnamefont{Hardy}}, \bibinfo {author}
  {\bibfnamefont{D.~A.}\ \bibnamefont{Bonn}}, \bibinfo {author}
  {\bibfnamefont{F.~F.}\ \bibnamefont{Balakirev}}, \bibinfo {author}
  {\bibfnamefont{R.~D.}\ \bibnamefont{McDonald}}, \bibinfo {author}
  {\bibfnamefont{J.~B.}\ \bibnamefont{Betts}}, \bibinfo {author}
  {\bibfnamefont{S.~C.}\ \bibnamefont{Riggs}},\ and\ \bibinfo {author}
  {\bibfnamefont{A.}~\bibnamefont{Migliori}},\ }%
  \bibfield{journal}{%
  \bibinfo {journal} {Nature}\ }%
  \textbf{\bibinfo {volume} {498}},\ \bibinfo {pages} {75} (\bibinfo {month}
  {06}\ \bibinfo {year} {2013}),\ \url{http://dx.doi.org/10.1038/nature12165}%
  \bibAnnoteFile{NoStop}{shekhter}%
\bibitem{lotfi}%
  \BibitemOpen
  \bibfield{author}{%
  \bibinfo {author} {\bibfnamefont{e.~a.}\ \bibnamefont{Z.~Lotfi}},\ }%
  \enquote{\bibinfo {title} {Universal inhomogeneous magnetic-field response in
  the normal state of cuprate high-tc superconductors},}\ %
  \bibAnnoteFile{NoStop}{lotfi}%
\bibitem{greven1}%
  \BibitemOpen
  \bibfield{author}{%
  \bibinfo {author} {\bibfnamefont{e.~a.}\ \bibnamefont{A.~m. Mounce}},\ }%
  \bibfield{journal}{%
  \bibinfo {journal} {http://arxiv.org/pdf/1304.6415v1.pdf}}%
   (\bibinfo {year} {2013})%
  \bibAnnoteFile{NoStop}{greven1}%
\bibitem{bourges}%
  \BibitemOpen
  \bibfield{author}{%
  \bibinfo {author} {\bibfnamefont{Y.}~\bibnamefont{Li}}, \bibinfo {author}
  {\bibfnamefont{V.}~\bibnamefont{Baledent}}, \bibinfo {author}
  {\bibfnamefont{G.}~\bibnamefont{Yu}}, \bibinfo {author}
  {\bibfnamefont{N.}~\bibnamefont{Barisic}}, \bibinfo {author}
  {\bibfnamefont{K.}~\bibnamefont{Hradil}}, \bibinfo {author}
  {\bibfnamefont{R.~A.}\ \bibnamefont{Mole}}, \bibinfo {author}
  {\bibfnamefont{Y.}~\bibnamefont{Sidis}}, \bibinfo {author}
  {\bibfnamefont{P.}~\bibnamefont{Steffens}}, \bibinfo {author}
  {\bibfnamefont{X.}~\bibnamefont{Zhao}}, \bibinfo {author}
  {\bibfnamefont{P.}~\bibnamefont{Bourges}},\ and\ \bibinfo {author}
  {\bibfnamefont{M.}~\bibnamefont{Greven}},\ }%
  \bibfield{journal}{%
  \bibinfo {journal} {Nature}\ }%
  \textbf{\bibinfo {volume} {468}},\ \bibinfo {pages} {283} (\bibinfo {month}
  {11}\ \bibinfo {year} {2010}),\ \url{http://dx.doi.org/10.1038/nature09477}%
  \bibAnnote{NoStop}{bourges}{10.1038/nature09477.}%
\bibitem{trsb1}%
  \BibitemOpen
  \bibfield{author}{%
  \bibinfo {author} {\bibfnamefont{A.}~\bibnamefont{Kaminski}}, \bibinfo
  {author} {\bibfnamefont{S.}~\bibnamefont{Rosenkranz}}, \bibinfo {author}
  {\bibfnamefont{H.~M.}\ \bibnamefont{Fretwell}}, \bibinfo {author}
  {\bibfnamefont{J.~C.}\ \bibnamefont{Campuzano}}, \bibinfo {author}
  {\bibfnamefont{Z.}~\bibnamefont{Li}}, \bibinfo {author}
  {\bibfnamefont{H.}~\bibnamefont{Raffy}}, \bibinfo {author}
  {\bibfnamefont{W.~G.}\ \bibnamefont{Cullen}}, \bibinfo {author}
  {\bibfnamefont{H.}~\bibnamefont{You}}, \bibinfo {author}
  {\bibfnamefont{C.~G.}\ \bibnamefont{Olson}}, \bibinfo {author}
  {\bibfnamefont{C.~M.}\ \bibnamefont{Varma}},\ and\ \bibinfo {author}
  {\bibfnamefont{H.}~\bibnamefont{Hochst}},\ }%
  \bibfield{journal}{%
  \bibinfo {journal} {Nature}\ }%
  \textbf{\bibinfo {volume} {416}},\ \bibinfo {pages} {610} (\bibinfo {month}
  {04}\ \bibinfo {year} {2002}),\ \url{http://dx.doi.org/10.1038/416610a}%
  \bibAnnote{NoStop}{trsb1}{10.1038/416610a.}%
\bibitem{trsb2}%
  \BibitemOpen
  \bibfield{author}{%
  \bibinfo {author} {\bibfnamefont{B.}~\bibnamefont{Fauqu\'e}}, \bibinfo
  {author} {\bibfnamefont{Y.}~\bibnamefont{Sidis}}, \bibinfo {author}
  {\bibfnamefont{V.}~\bibnamefont{Hinkov}}, \bibinfo {author}
  {\bibfnamefont{S.}~\bibnamefont{Pailh\`es}}, \bibinfo {author}
  {\bibfnamefont{C.~T.}\ \bibnamefont{Lin}}, \bibinfo {author}
  {\bibfnamefont{X.}~\bibnamefont{Chaud}},\ and\ \bibinfo {author}
  {\bibfnamefont{P.}~\bibnamefont{Bourges}},\ }%
  \bibfield{journal}{%
  \Doi{10.1103/PhysRevLett.96.197001}{\bibinfo {journal} {Phys. Rev. Lett.}}\
  }%
  \textbf{\bibinfo {volume} {96}},\ \bibinfo {pages} {197001} (\bibinfo {month}
  {May}\ \bibinfo {year} {2006}),\
  \url{http://link.aps.org/doi/10.1103/PhysRevLett.96.197001}%
  \bibAnnoteFile{NoStop}{trsb2}%
\bibitem{trsb3}%
  \BibitemOpen
  \bibfield{author}{%
  \bibinfo {author} {\bibfnamefont{J.}~\bibnamefont{Xia}}, \bibinfo {author}
  {\bibfnamefont{E.}~\bibnamefont{Schemm}}, \bibinfo {author}
  {\bibfnamefont{G.}~\bibnamefont{Deutscher}}, \bibinfo {author}
  {\bibfnamefont{S.~A.}\ \bibnamefont{Kivelson}}, \bibinfo {author}
  {\bibfnamefont{D.~A.}\ \bibnamefont{Bonn}}, \bibinfo {author}
  {\bibfnamefont{W.~N.}\ \bibnamefont{Hardy}}, \bibinfo {author}
  {\bibfnamefont{R.}~\bibnamefont{Liang}}, \bibinfo {author}
  {\bibfnamefont{W.}~\bibnamefont{Siemons}}, \bibinfo {author}
  {\bibfnamefont{G.}~\bibnamefont{Koster}}, \bibinfo {author}
  {\bibfnamefont{M.~M.}\ \bibnamefont{Fejer}},\ and\ \bibinfo {author}
  {\bibfnamefont{A.}~\bibnamefont{Kapitulnik}},\ }%
  \bibfield{journal}{%
  \Doi{10.1103/PhysRevLett.100.127002}{\bibinfo {journal} {Phys. Rev. Lett.}}\
  }%
  \textbf{\bibinfo {volume} {100}},\ \bibinfo {pages} {127002} (\bibinfo
  {month} {Mar}\ \bibinfo {year} {2008}),\
  \url{http://link.aps.org/doi/10.1103/PhysRevLett.100.127002}%
  \bibAnnoteFile{NoStop}{trsb3}%
\bibitem{trsb4}%
  \BibitemOpen
  \bibfield{author}{%
  \bibinfo {author} {\bibfnamefont{C.~M.}\ \bibnamefont{Varma}},\ }%
  \bibfield{journal}{%
  \Doi{10.1103/PhysRevB.73.155113}{\bibinfo {journal} {Phys. Rev. B}}\ }%
  \textbf{\bibinfo {volume} {73}},\ \bibinfo {pages} {155113} (\bibinfo {month}
  {Apr}\ \bibinfo {year} {2006}),\
  \url{http://link.aps.org/doi/10.1103/PhysRevB.73.155113}%
  \bibAnnoteFile{NoStop}{trsb4}%
\bibitem{louis}%
  \BibitemOpen
  \bibfield{author}{%
  \bibinfo {author} {\bibfnamefont{D.}~\bibnamefont{LeBoeuf}}, \bibinfo
  {author} {\bibfnamefont{N.}~\bibnamefont{Doiron-Leyraud}}, \bibinfo {author}
  {\bibfnamefont{B.}~\bibnamefont{Vignolle}}, \bibinfo {author}
  {\bibfnamefont{M.}~\bibnamefont{Sutherland}}, \bibinfo {author}
  {\bibfnamefont{B.~J.}\ \bibnamefont{Ramshaw}}, \bibinfo {author}
  {\bibfnamefont{J.}~\bibnamefont{Levallois}}, \bibinfo {author}
  {\bibfnamefont{R.}~\bibnamefont{Daou}}, \bibinfo {author}
  {\bibfnamefont{F.}~\bibnamefont{Lalibert\'e}}, \bibinfo {author}
  {\bibfnamefont{O.}~\bibnamefont{Cyr-Choini\`ere}}, \bibinfo {author}
  {\bibfnamefont{J.}~\bibnamefont{Chang}}, \bibinfo {author}
  {\bibfnamefont{Y.~J.}\ \bibnamefont{Jo}}, \bibinfo {author}
  {\bibfnamefont{L.}~\bibnamefont{Balicas}}, \bibinfo {author}
  {\bibfnamefont{R.}~\bibnamefont{Liang}}, \bibinfo {author}
  {\bibfnamefont{D.~A.}\ \bibnamefont{Bonn}}, \bibinfo {author}
  {\bibfnamefont{W.~N.}\ \bibnamefont{Hardy}}, \bibinfo {author}
  {\bibfnamefont{C.}~\bibnamefont{Proust}},\ and\ \bibinfo {author}
  {\bibfnamefont{L.}~\bibnamefont{Taillefer}},\ }%
  \bibfield{journal}{%
  \Doi{10.1103/PhysRevB.83.054506}{\bibinfo {journal} {Phys. Rev. B}}\ }%
  \textbf{\bibinfo {volume} {83}},\ \bibinfo {pages} {054506} (\bibinfo {month}
  {Feb}\ \bibinfo {year} {2011}),\
  \url{http://link.aps.org/doi/10.1103/PhysRevB.83.054506}%
  \bibAnnoteFile{NoStop}{louis}%
\bibitem{nernst}%
  \BibitemOpen
  \bibfield{author}{%
  \bibinfo {author} {\bibfnamefont{Z.~A.}\ \bibnamefont{Xu}}, \bibinfo {author}
  {\bibfnamefont{N.~P.}\ \bibnamefont{Ong}}, \bibinfo {author}
  {\bibfnamefont{Y.}~\bibnamefont{Wang}}, \bibinfo {author}
  {\bibfnamefont{T.}~\bibnamefont{Kakeshita}},\ and\ \bibinfo {author}
  {\bibfnamefont{S.}~\bibnamefont{Uchida}},\ }%
  \bibfield{journal}{%
  \bibinfo {journal} {Nature}\ }%
  \textbf{\bibinfo {volume} {406}},\ \bibinfo {pages} {486} (\bibinfo {month}
  {08}\ \bibinfo {year} {2000}),\ \url{http://dx.doi.org/10.1038/35020016}%
  \bibAnnote{NoStop}{nernst}{10.1038/35020016.}%
\bibitem{ando}%
  \BibitemOpen
  \bibfield{author}{%
  \bibinfo {author} {\bibfnamefont{Y.}~\bibnamefont{Ando}}, \bibinfo {author}
  {\bibfnamefont{S.}~\bibnamefont{Komiya}}, \bibinfo {author}
  {\bibfnamefont{K.}~\bibnamefont{Segawa}}, \bibinfo {author}
  {\bibfnamefont{S.}~\bibnamefont{Ono}},\ and\ \bibinfo {author}
  {\bibfnamefont{Y.}~\bibnamefont{Kurita}},\ }%
  \bibfield{journal}{%
  \Doi{10.1103/PhysRevLett.93.267001}{\bibinfo {journal} {Phys. Rev. Lett.}}\
  }%
  \textbf{\bibinfo {volume} {93}},\ \bibinfo {pages} {267001} (\bibinfo {month}
  {Dec}\ \bibinfo {year} {2004})%
  \bibAnnoteFile{NoStop}{ando}%
\bibitem{hussey}%
  \BibitemOpen
  \bibfield{author}{%
  \bibinfo {author} {\bibfnamefont{C.~R. A. X. X. W. Y. V.~B.}\
  \bibnamefont{Hussey}, \bibfnamefont{N.~E.}}\ and\ \bibinfo {author}
  {\bibfnamefont{C.}~\bibnamefont{Proust}},\ }%
  \bibfield{journal}{%
  \bibinfo {journal} {arXiv:0912.2001}}%
   (\bibinfo {year} {2009})%
  \bibAnnoteFile{NoStop}{hussey}%
\bibitem{greven}%
  \BibitemOpen
  \bibfield{author}{%
  \bibinfo {author} {\bibfnamefont{S.~I.}\ \bibnamefont{Mirzaei}}, \bibinfo
  {author} {\bibfnamefont{D.}~\bibnamefont{Stricker}}, \bibinfo {author}
  {\bibfnamefont{J.~N.}\ \bibnamefont{Hancock}}, \bibinfo {author}
  {\bibfnamefont{C.}~\bibnamefont{Berthod}}, \bibinfo {author}
  {\bibfnamefont{A.}~\bibnamefont{Georges}}, \bibinfo {author}
  {\bibfnamefont{E.}~\bibnamefont{van Heumen}}, \bibinfo {author}
  {\bibfnamefont{M.~K.}\ \bibnamefont{Chan}}, \bibinfo {author}
  {\bibfnamefont{X.}~\bibnamefont{Zhao}}, \bibinfo {author}
  {\bibfnamefont{Y.}~\bibnamefont{Li}}, \bibinfo {author}
  {\bibfnamefont{M.}~\bibnamefont{Greven}}, \bibinfo {author}
  {\bibfnamefont{N.}~\bibnamefont{Bari{\v s}i{\'c}}},\ and\ \bibinfo {author}
  {\bibfnamefont{D.}~\bibnamefont{van~der Marel}},\ }%
  \bibfield{journal}{%
  \Doi{10.1073/pnas.1218846110}{\bibinfo {journal} {Proceedings of the National
  Academy of Sciences}}\ }%
  \textbf{\bibinfo {volume} {110}},\ \bibinfo {pages} {5774} (\bibinfo {year}
  {2013}),\
  \Eprint{http://arxiv.org/abs/http://www.pnas.org/content/110/15/5774.full.pd%
f+html}{http://www.pnas.org/content/110/15/5774.full.pdf+html},\
  \url{http://www.pnas.org/content/110/15/5774.abstract}%
  \bibAnnoteFile{NoStop}{greven}%
\bibitem{mfl}%
  \BibitemOpen
  \bibfield{author}{%
  \bibinfo {author} {\bibfnamefont{C.~M.}\ \bibnamefont{Varma}}, \bibinfo
  {author} {\bibfnamefont{P.~B.}\ \bibnamefont{Littlewood}}, \bibinfo {author}
  {\bibfnamefont{S.}~\bibnamefont{Schmitt-Rink}}, \bibinfo {author}
  {\bibfnamefont{E.}~\bibnamefont{Abrahams}},\ and\ \bibinfo {author}
  {\bibfnamefont{A.~E.}\ \bibnamefont{Ruckenstein}},\ }%
  \bibfield{journal}{%
  \bibinfo {journal} {Physical Review Letters}\ }%
  \textbf{\bibinfo {volume} {63}} (\bibinfo {year} {1989}),\
  \url{http://link.aps.org/abstract/PRL/v63/p1996}%
  \bibAnnoteFile{NoStop}{mfl}%
\bibitem{yrz}%
  \BibitemOpen
  \bibfield{author}{%
  \bibinfo {author} {\bibfnamefont{K.-Y.}\ \bibnamefont{Yang}}, \bibinfo
  {author} {\bibfnamefont{T.~M.}\ \bibnamefont{Rice}},\ and\ \bibinfo {author}
  {\bibfnamefont{F.-C.}\ \bibnamefont{Zhang}},\ }%
  \bibfield{journal}{%
  \Doi{10.1103/PhysRevB.73.174501}{\bibinfo {journal} {Phys. Rev. B}}\ }%
  \textbf{\bibinfo {volume} {73}},\ \bibinfo {pages} {174501} (\bibinfo {month}
  {May}\ \bibinfo {year} {2006}),\
  \url{http://link.aps.org/doi/10.1103/PhysRevB.73.174501}%
  \bibAnnoteFile{NoStop}{yrz}%
\bibitem{ftm1}%
  \BibitemOpen
  \bibfield{author}{%
  \bibinfo {author} {\bibfnamefont{R.~G.}\ \bibnamefont{Leigh}}, \bibinfo
  {author} {\bibfnamefont{P.}~\bibnamefont{Phillips}},\ and\ \bibinfo {author}
  {\bibfnamefont{T.-P.}\ \bibnamefont{Choy}},\ }%
  \bibfield{journal}{%
  \bibinfo {journal} {Phys. Rev. Lett.}\ }%
  \textbf{\bibinfo {volume} {99}},\ \bibinfo {pages} {046404} (\bibinfo {year}
  {2007})%
  \bibAnnoteFile{NoStop}{ftm1}%
\bibitem{ftm2}%
  \BibitemOpen
  \bibfield{author}{%
  \bibinfo {author} {\bibfnamefont{T.-P.}\ \bibnamefont{Choy}}, \bibinfo
  {author} {\bibfnamefont{R.~G.}\ \bibnamefont{Leigh}}, \bibinfo {author}
  {\bibfnamefont{P.}~\bibnamefont{Phillips}},\ and\ \bibinfo {author}
  {\bibfnamefont{P.~D.}\ \bibnamefont{Powell}},\ }%
  \bibfield{journal}{%
  \bibinfo {journal} {Phys. Rev. B}\ }%
  \textbf{\bibinfo {volume} {77}},\ \bibinfo {pages} {014512} (\bibinfo {year}
  {2008})%
  \bibAnnoteFile{NoStop}{ftm2}%
\bibitem{ftm3}%
  \BibitemOpen
  \bibfield{author}{%
  \bibinfo {author} {\bibfnamefont{P.}~\bibnamefont{Phillips}},\ }%
  \bibfield{journal}{%
  \Doi{10.1103/RevModPhys.82.1719}{\bibinfo {journal} {Rev. Mod. Phys.}}\ }%
  \textbf{\bibinfo {volume} {82}},\ \bibinfo {pages} {1719} (\bibinfo {month}
  {May}\ \bibinfo {year} {2010}),\
  \url{http://link.aps.org/doi/10.1103/RevModPhys.82.1719}%
  \bibAnnoteFile{NoStop}{ftm3}%
\bibitem{Meir1992}%
  \BibitemOpen
  \bibfield{author}{%
  \bibinfo {author} {\bibfnamefont{Y.}~\bibnamefont{Meir}}\ and\ \bibinfo
  {author} {\bibfnamefont{N.~S.}\ \bibnamefont{Wingreen}},\ }%
  \bibfield{journal}{%
  \Doi{10.1103/PhysRevLett.68.2512}{\bibinfo {journal} {Phys. Rev. Lett.}}\ }%
  \textbf{\bibinfo {volume} {68}},\ \bibinfo {pages} {2512} (\bibinfo {month}
  {Apr}\ \bibinfo {year} {1992}),\
  \url{http://link.aps.org/doi/10.1103/PhysRevLett.68.2512}%
  \bibAnnoteFile{NoStop}{Meir1992}%
\bibitem{Jauho1994}%
  \BibitemOpen
  \bibfield{author}{%
  \bibinfo {author} {\bibfnamefont{A.-P.}\ \bibnamefont{Jauho}}, \bibinfo
  {author} {\bibfnamefont{N.~S.}\ \bibnamefont{Wingreen}},\ and\ \bibinfo
  {author} {\bibfnamefont{Y.}~\bibnamefont{Meir}},\ }%
  \bibfield{journal}{%
  \Doi{10.1103/PhysRevB.50.5528}{\bibinfo {journal} {Phys. Rev. B}}\ }%
  \textbf{\bibinfo {volume} {50}},\ \bibinfo {pages} {5528} (\bibinfo {month}
  {Aug}\ \bibinfo {year} {1994}),\
  \url{http://link.aps.org/doi/10.1103/PhysRevB.50.5528}%
  \bibAnnoteFile{NoStop}{Jauho1994}%
\bibitem{schwinger}%
  \BibitemOpen
  \bibfield{author}{%
  \bibinfo {author} {\bibfnamefont{J.}~\bibnamefont{Schwinger}},\ }%
  \bibfield{journal}{%
  \bibinfo {journal} {J. Math Phys. (N. Y.)}\ }%
  \textbf{\bibinfo {volume} {2}} (\bibinfo {year} {1961})%
  \bibAnnoteFile{NoStop}{schwinger}%
\bibitem{baym}%
  \BibitemOpen
  \bibfield{author}{%
  \bibinfo {author} {\bibfnamefont{L.~P.}\ \bibnamefont{Kadanoff}}\ and\
  \bibinfo {author} {\bibfnamefont{G.}~\bibnamefont{Baym}},\ }%
  \emph{\bibinfo {title} {Quantum Statistical Mechanics}}\ (\bibinfo
  {publisher} {Benjamin, New York},\ \bibinfo {year} {1962})%
  \bibAnnoteFile{NoStop}{baym}%
\bibitem{Keldysh1965}%
  \BibitemOpen
  \bibfield{author}{%
  \bibinfo {author} {\bibfnamefont{L.~V.}\ \bibnamefont{Keldysh}},\ }%
  \bibfield{journal}{%
  \bibinfo {journal} {Zh. Eksp. Teor. Fiz.}\ }%
  \textbf{\bibinfo {volume} {47}} (\bibinfo {year} {1965})%
  \bibAnnoteFile{NoStop}{Keldysh1965}%
\bibitem{shong}%
  \BibitemOpen
  \bibfield{author}{%
  \bibinfo {author} {\bibfnamefont{S.}~\bibnamefont{Hong}}\ and\ \bibinfo
  {author} {\bibfnamefont{P.}~\bibnamefont{Phillips}},\ }%
  \bibfield{journal}{%
  \Doi{10.1103/PhysRevB.86.115118}{\bibinfo {journal} {Phys. Rev. B}}\ }%
  \textbf{\bibinfo {volume} {86}},\ \bibinfo {pages} {115118} (\bibinfo {month}
  {Sep}\ \bibinfo {year} {2012}),\
  \url{http://link.aps.org/doi/10.1103/PhysRevB.86.115118}%
  \bibAnnoteFile{NoStop}{shong}%
\bibitem{morr01}%
  \BibitemOpen
  \bibfield{author}{%
  \bibinfo {author} {\bibfnamefont{S.}~\bibnamefont{Chakravarty}}, \bibinfo
  {author} {\bibfnamefont{R.~B.}\ \bibnamefont{Laughlin}}, \bibinfo {author}
  {\bibfnamefont{D.~K.}\ \bibnamefont{Morr}},\ and\ \bibinfo {author}
  {\bibfnamefont{C.}~\bibnamefont{Nayak}},\ }%
  \bibfield{journal}{%
  \Doi{10.1103/PhysRevB.63.094503}{\bibinfo {journal} {Phys. Rev. B}}\ }%
  \textbf{\bibinfo {volume} {63}},\ \bibinfo {pages} {094503} (\bibinfo {month}
  {Jan}\ \bibinfo {year} {2001}),\
  \url{http://link.aps.org/doi/10.1103/PhysRevB.63.094503}%
  \bibAnnoteFile{NoStop}{morr01}%
\bibitem{Norman2007}%
  \BibitemOpen
  \bibfield{author}{%
  \bibinfo {author} {\bibfnamefont{M.~R.}\ \bibnamefont{Norman}}, \bibinfo
  {author} {\bibfnamefont{A.}~\bibnamefont{Kanigel}}, \bibinfo {author}
  {\bibfnamefont{M.}~\bibnamefont{Randeria}}, \bibinfo {author}
  {\bibfnamefont{U.}~\bibnamefont{Chatterjee}},\ and\ \bibinfo {author}
  {\bibfnamefont{J.~C.}\ \bibnamefont{Campuzano}},\ }%
  \bibfield{journal}{%
  \Doi{10.1103/PhysRevB.76.174501}{\bibinfo {journal} {Phys. Rev. B}}\ }%
  \textbf{\bibinfo {volume} {76}},\ \bibinfo {pages} {174501} (\bibinfo {month}
  {Nov}\ \bibinfo {year} {2007}),\
  \url{http://link.aps.org/doi/10.1103/PhysRevB.76.174501}%
  \bibAnnoteFile{NoStop}{Norman2007}%
\bibitem{moon2010}%
  \BibitemOpen
  \bibfield{author}{%
  \bibinfo {author} {\bibfnamefont{E.~G.}\ \bibnamefont{Moon}}\ and\ \bibinfo
  {author} {\bibfnamefont{S.}~\bibnamefont{Sachdev}},\ }%
  \bibfield{journal}{%
  \Doi{10.1103/PhysRevB.80.035117}{\bibinfo {journal} {Phys. Rev. B}}\ }%
  \textbf{\bibinfo {volume} {80}},\ \bibinfo {pages} {035117} (\bibinfo {month}
  {Jul}\ \bibinfo {year} {2009}),\
  \url{http://link.aps.org/doi/10.1103/PhysRevB.80.035117}%
  \bibAnnoteFile{NoStop}{moon2010}%
\bibitem{morr98}%
  \BibitemOpen
  \bibfield{author}{%
  \bibinfo {author} {\bibnamefont{{B. L. Altshuler}}}, \bibinfo {author}
  {\bibnamefont{{A. V. Chubukov}}}, \bibinfo {author} {\bibnamefont{{A.
  Dashevskii}}}, \bibinfo {author} {\bibnamefont{{A. M. Finkel'stein}}},\ and\
  \bibinfo {author} {\bibnamefont{{D. K. Morr}}},\ }%
  \bibfield{journal}{%
  \Doi{10.1209/epl/i1998-00164-y}{\bibinfo {journal} {Europhys. Lett.}}\ }%
  \textbf{\bibinfo {volume} {41}},\ \bibinfo {pages} {401} (\bibinfo {year}
  {1998}),\ \url{http://dx.doi.org/10.1209/epl/i1998-00164-y}%
  \bibAnnoteFile{NoStop}{morr98}%
\bibitem{dave}%
  \BibitemOpen
  \bibfield{author}{%
  \bibinfo {author} {\bibfnamefont{C.~L.~K.}\ \bibnamefont{K.~Dave},
  \bibfnamefont{P.~W.~Phillips}},\ }%
  \bibfield{journal}{%
  \bibinfo {journal} {http://arxiv.org/abs/1207.4201}}%
   (\bibinfo {year} {2012})%
  \bibAnnoteFile{NoStop}{dave}%
\bibitem{kotstan}%
  \BibitemOpen
  \bibfield{author}{%
  \bibinfo {author} {\bibfnamefont{T.~D.}\ \bibnamefont{Stanescu}}\ and\
  \bibinfo {author} {\bibfnamefont{G.}~\bibnamefont{Kotliar}},\ }%
  \bibfield{journal}{%
  \Doi{10.1103/PhysRevB.74.125110}{\bibinfo {journal} {Phys. Rev. B}}\ }%
  \textbf{\bibinfo {volume} {74}},\ \bibinfo {pages} {125110} (\bibinfo {month}
  {Sep}\ \bibinfo {year} {2006}),\
  \url{http://link.aps.org/doi/10.1103/PhysRevB.74.125110}%
  \bibAnnoteFile{NoStop}{kotstan}%
\bibitem{civelli11}%
  \BibitemOpen
  \bibfield{author}{%
  \bibinfo {author} {\bibfnamefont{S.}~\bibnamefont{Sakai}}, \bibinfo {author}
  {\bibfnamefont{G.}~\bibnamefont{Sangiovanni}}, \bibinfo {author}
  {\bibfnamefont{M.}~\bibnamefont{Civelli}}, \bibinfo {author}
  {\bibfnamefont{Y.}~\bibnamefont{Motome}}, \bibinfo {author}
  {\bibfnamefont{K.}~\bibnamefont{Held}},\ and\ \bibinfo {author}
  {\bibfnamefont{M.}~\bibnamefont{Imada}},\ }%
  \bibfield{journal}{%
  \Doi{10.1103/PhysRevB.85.035102}{\bibinfo {journal} {Phys. Rev. B}}\ }%
  \textbf{\bibinfo {volume} {85}},\ \bibinfo {pages} {035102} (\bibinfo {month}
  {Jan}\ \bibinfo {year} {2012}),\
  \url{http://link.aps.org/doi/10.1103/PhysRevB.85.035102}%
  \bibAnnoteFile{NoStop}{civelli11}%
\bibitem{tremblay}%
  \BibitemOpen
  \bibfield{author}{%
  \bibinfo {author} {\bibfnamefont{A.~M.}\ \bibnamefont{Tremblay}}}%
   (\bibinfo {month} {September}\ \bibinfo {year} {2010})%
  \bibAnnoteFile{NoStop}{tremblay}%
\bibitem{Yonemitsu2009}%
  \BibitemOpen
  \bibfield{author}{%
  \bibinfo {author} {\bibfnamefont{K.}~\bibnamefont{Yonemitsu}},\ }%
  \bibfield{journal}{%
  \Doi{10.1143/JPSJ.78.054705}{\bibinfo {journal} {Journal of the Physical
  Society of Japan}}\ }%
  \textbf{\bibinfo {volume} {78}},\ \bibinfo {pages} {054705} (\bibinfo {year}
  {2009}),\ \url{http://jpsj.ipap.jp/link?JPSJ/78/054705/}%
  \bibAnnoteFile{NoStop}{Yonemitsu2009}%
\bibitem{Zhang1988}%
  \BibitemOpen
  \bibfield{author}{%
  \bibinfo {author} {\bibfnamefont{F.~C.}\ \bibnamefont{Zhang}}, \bibinfo
  {author} {\bibfnamefont{C.}~\bibnamefont{Gros}}, \bibinfo {author}
  {\bibfnamefont{T.~M.}\ \bibnamefont{Rice}},\ and\ \bibinfo {author}
  {\bibfnamefont{H.}~\bibnamefont{Shiba}},\ }%
  \bibfield{journal}{%
  \bibinfo {journal} {Superconductor Science and Technology}\ }%
  \textbf{\bibinfo {volume} {1}},\ \bibinfo {pages} {36} (\bibinfo {year}
  {1988}),\ \url{http://stacks.iop.org/0953-2048/1/i=1/a=009}%
  \bibAnnoteFile{NoStop}{Zhang1988}%
\bibitem{Ng1996}%
  \BibitemOpen
  \bibfield{author}{%
  \bibinfo {author} {\bibfnamefont{T.-K.}\ \bibnamefont{Ng}},\ }%
  \bibfield{journal}{%
  \Doi{10.1103/PhysRevLett.76.487}{\bibinfo {journal} {Phys. Rev. Lett.}}\ }%
  \textbf{\bibinfo {volume} {76}},\ \bibinfo {pages} {487} (\bibinfo {month}
  {Jan}\ \bibinfo {year} {1996}),\
  \url{http://link.aps.org/doi/10.1103/PhysRevLett.76.487}%
  \bibAnnoteFile{NoStop}{Ng1996}%
\bibitem{imada}%
  \BibitemOpen
  \bibfield{author}{%
  \bibinfo {author} {\bibfnamefont{Y.}~\bibnamefont{Yamaji}}\ and\ \bibinfo
  {author} {\bibfnamefont{M.}~\bibnamefont{Imada}},\ }%
  \bibfield{journal}{%
  \Doi{10.1103/PhysRevLett.106.016404}{\bibinfo {journal} {Phys. Rev. Lett.}}\
  }%
  \textbf{\bibinfo {volume} {106}},\ \bibinfo {pages} {016404} (\bibinfo
  {month} {Jan}\ \bibinfo {year} {2011}),\
  \url{http://link.aps.org/doi/10.1103/PhysRevLett.106.016404}%
  \bibAnnoteFile{NoStop}{imada}%
\bibitem{Yang2010}%
  \BibitemOpen
  \bibfield{author}{%
  \bibinfo {author} {\bibfnamefont{K.-Y.}\ \bibnamefont{Yang}}, \bibinfo
  {author} {\bibfnamefont{K.}~\bibnamefont{Huang}}, \bibinfo {author}
  {\bibfnamefont{W.-Q.}\ \bibnamefont{Chen}}, \bibinfo {author}
  {\bibfnamefont{T.~M.}\ \bibnamefont{Rice}},\ and\ \bibinfo {author}
  {\bibfnamefont{F.-C.}\ \bibnamefont{Zhang}},\ }%
  \bibfield{journal}{%
  \Doi{10.1103/PhysRevLett.105.167004}{\bibinfo {journal} {Phys. Rev. Lett.}}\
  }%
  \textbf{\bibinfo {volume} {105}},\ \bibinfo {pages} {167004} (\bibinfo
  {month} {Oct}\ \bibinfo {year} {2010}),\
  \url{http://link.aps.org/doi/10.1103/PhysRevLett.105.167004}%
  \bibAnnoteFile{NoStop}{Yang2010}%
\bibitem{sawatzky}%
  \BibitemOpen
  \bibfield{author}{%
  \bibinfo {author} {\bibfnamefont{M.~B.~J.}\ \bibnamefont{Meinders}}, \bibinfo
  {author} {\bibfnamefont{H.}~\bibnamefont{Eskes}},\ and\ \bibinfo {author}
  {\bibfnamefont{G.~A.}\ \bibnamefont{Sawatzky}},\ }%
  \bibfield{journal}{%
  \Doi{10.1103/PhysRevB.48.3916}{\bibinfo {journal} {Phys. Rev. B}}\ }%
  \textbf{\bibinfo {volume} {48}},\ \bibinfo {pages} {3916} (\bibinfo {month}
  {Aug}\ \bibinfo {year} {1993})%
  \bibAnnoteFile{NoStop}{sawatzky}%
\end{thebibliography}
%
\end{document}